\documentclass[useAMS]{mn2e}
\usepackage{times}
\input{epsf}

\title[Energy-dependent variability and soft excess]{Energy-dependent variability and
the origin of the soft X-ray excess in AGN}

\author[M.~Gierli{\'n}ski and C.~Done] {Marek Gierli\'{n}ski$^{1,2}$ and Chris Done$^1$\\
$^1$Department of Physics, University of Durham, South Road, Durham DH1 3LE,UK\\
$^2$Obserwatorium Astronomiczne Uniwersytetu Jagiello\'{n}skiego,
30-244 Krak\'{o}w, Orla 171, Poland}

\date{Submitted to MNRAS}
\pagerange{\pageref{firstpage}--\pageref{lastpage}}

\begin{document}

\maketitle

\label{firstpage}

\begin{abstract}

The origin of the soft excess seen in many AGN below $\sim$1~keV is
still an unsolved problem. It is unlikely to represent a true continuum
component as its characteristic `temperature' shows a remarkable
constancy over a wide range of AGN luminosity and black hole mass. This
instead favours an association with atomic processes, in particular
with the increase in opacity between 0.7--2~keV associated with
partially ionized O and Fe. The opacity jump can give rise to a soft
excess either through reflection or transmission, and both scenarios
can fit the spectra equally well as long as there is strong velocity
shear to smear out the characteristic narrow atomic features. Here we
use orthogonal constraints from the energy-dependent variability. The
rms spectra seen in {\it XMM-Newton} AGN data often show a broad peak
between 0.7--2~keV. We show that the absorption model can explain the
data well if the ionization state of the smeared absorption responds to
luminosity changes in the continuum.

\end{abstract}

\begin{keywords}
  accretion, accretion discs -- atomic processes -- galaxies: active -- X-rays:
  galaxies.
\end{keywords}

\section{Introduction}

One of the most interesting problems in AGN astrophysics is the origin
of the soft X-ray excess in Seyfert 1s and radio-quiet quasars. The
soft X-ray spectra of these objects often show extra emission below
$\sim$1 keV with respect to the extrapolated power law observed at
higher energies. This was originally interpreted as thermal emission
from the (Comptonized) accretion disc (Turner \& Pounds 1988; Magdziarz
et al. 1998). However, the temperature of this component is remarkably
constant around 0.1--0.2~keV regardless of central object luminosity
and mass, in conflict with disc models (Czerny et al 2003;
Gierli{\'n}ski \& Done 2004, hereafter GD04; Crummy et al. 2006).

The constant temperature of the soft excess is very difficult to
reconcile within any model of continuum emission.  It is much easier
to explain if it has an origin in atomic processes, as these have
characteristic energies. In particular, there is a strong jump in
opacity at $\sim$0.7 keV from partially ionized material, where
O\,{\sc vii}/O\,{\sc viii} and the Fe M shell unresolved transition
array (UTA) combine to produce much more absorption above this energy
than below.  However, the soft excess is very smooth, with no
discernable line and edge features expected from atomic processes.
Thus if this is the origin of the soft excess, the partially ionized
material must have strong velocity gradients so that Doppler shifts
smear these out.

Two alternative scenarios have been proposed using this underlying
physics. The first is that the partially ionized, velocity smeared
material is seen in reflection (e.g. Fabian et al. 2002a, 2004, 2005;
Crummy et al. 2006), the other that it is seen in absorption (GD04;
Chevallier et al. 2006; Schurch \& Done in press). The reflection
geometry sets a limit on the size of the soft X-ray excess to only a
factor $\sim$2--3 above the extrapolation of the higher energy spectrum
(assuming an isotropic source above a smooth disc surface), as the
maximum reflected flux is set by the level of the illuminating flux.
However, some sources are observed to have soft excesses which are much
stronger than this, leading to the additional requirement of
anisotropic illumination, perhaps from gravitational light bending,
and/or a corrugated disc surface which hides most of the intrinsic flux
from sight (Fabian et al. 2002a; Miniutti et al. 2003; Miniutti \&
Fabian 2004). The alternative geometry, where the partially ionized,
velocity smeared material is seen in absorption, removes this
requirement as it has no corresponding limit on the size of the soft
excess. Instead, it requires fast moving ionized material in the line
of sight (e.g. Sobolewska \& Done in preparation, hereafter SD06).

These two models represent rather different physical geometries, and
give rather different spectral deconvolutions of the data. However,
they are indistinguishable in terms of statistical fit quality even
with excellent spectral data constraining the shape of the soft excess
from {\it XMM-Newton} (Chevallier et al. 2006; SD06). This is
especially important as the interpretation of the soft excess impacts
on the inferred properties of the spacetime.  A reflection dominated
approach leads to inferred high-to-maximal black hole spin, and perhaps
also to an emissivity which is enhanced by extraction of the spin
energy of the black hole  (Wilms et al. 2001). This is seen both in
terms of the iron line smearing, and in the size of the soft excess
assuming that light bending is the origin of the anisotropic
illumination of the disc (Fabian et al. 2004; Crummy et al. 2006). By
contrast, with the absorption approach, SD06 and Schurch \& Done (in
press) show that this continuum curvature removes the requirement for
an extreme red wing to the iron line in PG 1211+143. There is still a
reflection component from the X-ray illuminated disc, but this does not
dominate the spectrum and the amount of relativistic smearing is
compatible with any black hole spin.

The idea that the extremely broad residual between 3 and 6 keV seen,
e.g., in MCG--6-30-15 could be an artifact of absorption induced
continuum curvature was originally proposed by Inoue \& Matsumoto
(2003) and Kinkhabwala (2003). They showed that absorption can indeed
match the shape of the broad residual seen in moderate resolution {\it
ASCA} data. However, Young et al. (2005) pointed out that this model
also predicts strong, narrow iron line absorption in the 6.45--6.5~keV
range, which is not seen in high resolution Chandra grating data. The
model of GD04 differs from this in that the absorption is smeared due
to a large (relativistic) range of velocities, so there are no narrow
absorption features, consistent with the Chandra limits.

Thus distinguishing between the absorption and reflection geometries is
very important in terms of understanding the spacetime of the black
hole as well as understanding the nature of the accretion flow. Since
the spectral approach is currently unable to discriminate between these
two models, we here investigate their variability properties in the
hope that this would give orthogonal constraints. Inoue \& Matsumoto
(2003) showed that the {\it ASCA} variability spectrum of MCG--6-30-15
is consistent with a varying power law and varying warm absorbers.
Here, we study in details the energy-dependent variability of an
ionized absorber due changes in the luminosity of the continuum and
show that this can match the observed variability spectra of three AGN.

\begin{figure}
\begin{center}
\leavevmode \epsfxsize=8cm \epsfbox{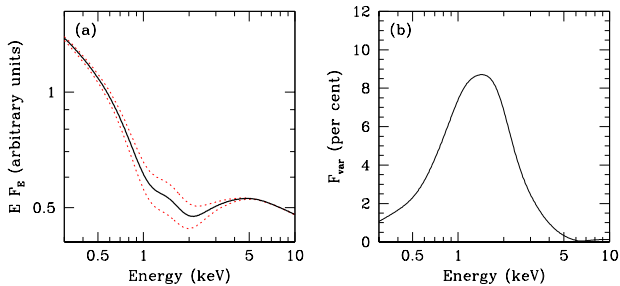}
\end{center}
\caption{The variability predicted by ionization changes in the smeared
absorption model for a column of $N_H$ = 10$^{23}$ cm$^{-2}$ and
Gaussian velocity smearing of $\sigma_{\rm vel}$ = 0.3 on an underlying
continuum with $\Gamma=2.3$. The ionization varies with amplitude
$r(\lg \xi) \equiv \sigma(\lg \xi) / \langle \lg \xi \rangle$ = 0.03
around the mean $\langle \lg \xi \rangle$ = 2.8. Panel (a): the energy
spectra corresponding to the mean ionization (solid line) and $\pm
\sigma(\lg \xi)$ (dotted lines); (b) The corresponding rms variability
spectrum. } \label{fig:varrms}
\end{figure}

\section{Energy-dependent variability}
\label{sec:rms}

\subsection{Models}
\label{sec:models}

The standard approach to investigate the energy dependence of the
variability is to plot the fractional root mean square (rms)
variability amplitude, $F_{\rm var}$, as a function of energy. The
techniques for calculating such spectra are described in Edelson et al.
(2002), Markowitz, Edelson \& Vaughan (2003) and Vaughan et al. (2003),
but in outline the method consists of extracting a background
subtracted light curve in each energy band, then calculating the
fractional rms variability in the light curve, and then correcting this
for the additional variance imposed by the measurement errors. These
techniques have been applied to data from several instruments covering
a range of energies: {\it RXTE} (3--20~keV), {\it ASCA} (0.7--10~keV)
and {\it XMM-Newton} (0.3--10~keV).

The rms spectrum from a given object need not to be unique. It depends
on the time-scales on which the data are measured, as these probe
different size-scales. {\it RXTE} monitored various AGN on all
time-scales between days and years (Markowitz et al. 2003), while {\it
ASCA} and {\it XMM-Newton} observed over much shorter time-scales
(hundreds of seconds to days). AGN variability can be also
non-stationary (Green, M$^c$Hardy \& Done 1999), so the rms spectra can
change in time. This was demonstrated by Gallo et al. (2004c) who
obtained dramatically different rms spectra depending on whether a
strong flare in the light curve was included or not.

Typically the rms spectra are consistent with a broad peak in the
variability amplitude between 0.8--2~keV, decreasing to higher and
lower energies, reaching a minimum around the Fe K$\alpha$ line, and
then increasing slightly above this energy (Fabian et al. 2002b; Inoue
\& Matsumoto 2003; Markowitz et al. 2003; Gallo et al. 2004a, b; Ponti
et al. 2004, 2006). This alone is enough to strongly support atomic
models for the origin of the soft excess, as it is very hard to imagine
the variability of any true soft continuum component producing such
energy dependance.

\begin{figure*}
\begin{center}
\leavevmode \epsfxsize=12cm \epsfbox{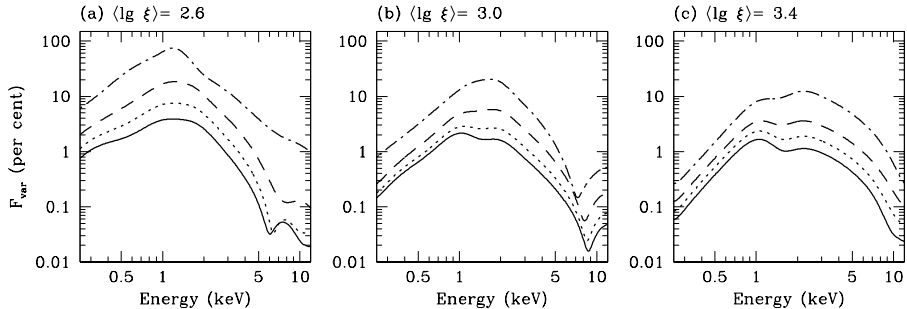}
\end{center}
\caption{Grid of variability models based on the smeared absorption,
with $\lg \xi$ varying around its mean value with the standard
deviation of 3 per cent of the mean. Each panel shows the series of
model variability spectra for $N_H$ = 3 (solid curve), 5 (dotted), 10
(dashed) and 30 (dash-dotted) $\times10^{22}$ cm$^{-2}$, with mean
$\langle \lg \xi \rangle$ = 2.6, 3.0 and 3.4 from left to right.}
\label{fig:grid}
\end{figure*}

A successful model should simultaneously be able to explain both the
spectrum and the energy-dependent variability. However, in general, the
X-ray spectra of AGN are complex and consist of many components, e.g.
cold absorption, warm absorbers, continuum and reflection (see, e.g.,
Fabian \& Miniutti in press). Here we first examined the variability
produced by the smeared absorption model of GD04. This is built around
the absorption calculated from the sophisticated photoionization code,
{\sc xstar} (Bautista \& Kallman 2001), and then smeared using a
Gaussian kernel. This approximates the behaviour of a large velocity
shear in the absorber, and has the effect of removing the
characteristic narrow absorption features, so as to produce a smooth
soft excess.

Three basic types of variability can be produced from such a model.
Firstly, the ionization state of the absorber, $\xi=L/nR^2$, (where
$L$, $n$ and $R$ are the luminosity, density and distance of the cloud
from the ionizing source, respectively) could change, in response to
the illuminating continuum variability. Secondly, its column density,
$N_H$, could change as a result of clouds moving in and out of the line
of sight. Finally, the velocity shear, parameterized by the width of
the smearing Gaussian, $\sigma_{\rm vel}$, could change. The most
physically plausible seems to be changes in ionization, especially as
it is clear that the continuum does vary (regardless whether it is due
to accretion rate change, or, e.g., light bending effects; Miniutti et
al. 2003). Hence, we first examined the effect of changing $\xi$.

We employed a Monte Carlo approach to quantify the spectral
variability. We simulated 3000 model spectra where $\lg
\xi$\footnote{We denote $\lg \xi \equiv \lg (\xi / $1 erg cm s$^{-1}$),
for the sake of clarity.} was varied randomly with Gaussian
distribution of standard deviation $\sigma(\lg \xi)$ [or fractional
rms, $r(\lg \xi) \equiv \sigma(\lg \xi) / \langle \lg \xi \rangle$]
around a given mean, $\langle \lg \xi \rangle$. All other parameters
were fixed. These model spectra were used to calculate the standard
deviation at each energy channel, and thus to produce the rms spectrum
(Gierli{\'n}ski \& Zdziarski 2005).

We started from parameters of the smeared absorber which are close to
those which fit the spectrum of PG 1211+143 (GD04; SD06; Schurch \&
Done in press): $N_H = 10^{23}$ cm$^{-2}$, $\langle \lg \xi \rangle
=2.8$ and $\sigma_{\rm vel} = 0.3$. The continuum was a power law with
the photon spectral index $\Gamma$ = 2.3. Such a soft intrinsic index
is the effect of the absorber hardening the apparent spectrum above
$\sim$1 keV. For example, SD06 show that the spectral index of the
underlying continuum in 1H 0707--495 is $\sim$2.5 in the smeared
absorber model, while the apparent 2--10 keV index is $\sim$1.8. We
point out, that the exact value of the spectral index has little effect
on our results.

Fig. \ref{fig:varrms} shows the effect of varying $\lg\xi$ by 3 per
cent of its mean value, i.e. $r(\lg \xi)$ = 0.03. Panel (a) shows the
spectral effect of this broad absorption. The two major sources of
opacity are at $\sim$0.9~keV (predominantly from O\,{\sc vii}/O\,{\sc
viii} lines and edges, together with the Fe UTA) and at 1.2--2~keV
(predominantly from Fe L lines and edges). At lower ionization the
absorption becomes more prominent, while at higher ionization the
medium becomes more transparent, but since there is little opacity
below 0.5~keV and above 5~keV, the variability spectrum (shown in
Fig.\ref{fig:varrms}b) results in a broad peak between these energies,
as is typically seen from AGN rms spectra (Fabian et al. 2002b;
Markowitz et al. 2003; Gallo et al. 2004a, b; Ponti et al. 2004, 2006).

We explored the range of variability spectra produced by different
columns and mean ionization parameters. The rms spectra shown in
Fig.~\ref{fig:grid} were calculated from 3 per cent variability in $\lg
\xi$, i.e. corresponding to changes of $\sim$20 per cent in $\xi$. The
intrinsic spectral index of the continuum was 2.3. All these show the
characteristic peak in the rms at 1--2~keV, but with very different
normalization. This is because the depth of the absorption trough in
the spectrum is a highly non-linear function of the ionization state. A
small decrease in $\xi$ leads to much larger increase in opacity for
$\langle \lg \xi \rangle=2.6$ than for $\langle \lg\xi \rangle=3.4$, so
giving much higher amplitude of variability for the same column at
lower mean ionization states. The behaviour with column is somewhat
simpler as it is not a function of ionization state. At lower columns
the absorption has a smaller effect on the spectrum, so the changes in
ionization give rise to less dramatic variability.

\begin{figure}
\begin{center}
\leavevmode \epsfxsize=8cm \epsfbox{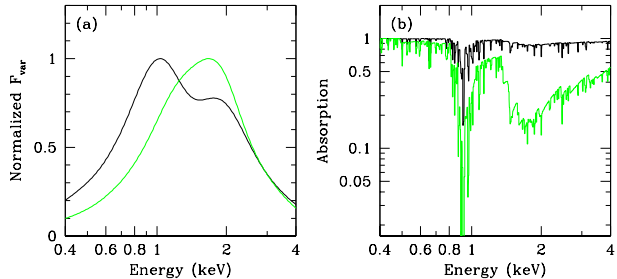}
\end{center}
\caption{(a) Detailed shape of rms spectra for $N_H$ = 3 (black) and 30
(grey, green in colour) $\times 10^{22}$ cm$^{-2}$ for $\langle \lg \xi
\rangle$ = 3.0 taken from Fig. \ref{fig:grid}(b), and normalized to
their peak values. (b) Underlying, unsmeared, mean absorption
corresponding to each column. The lower energy absorption system at
$\sim$0.9~keV consists mainly of resonance lines. These are saturated
in the high-column system, so small changes in ionization do not
produce much variability, and there is little contribution to the rms
(grey or green curve). Conversely, the lines are not saturated at lower
columns so this feature is prominent in the rms spectrum (black curve).
The higher energy absorption at 1.2--2~keV is dominated by iron L
edges. These do not saturate so this feature is present in both the
high and low column rms spectra. } \label{fig:pabs}
\end{figure}

There are also some detailed changes in the shape of the rms spectra.
We illustrate this by rescaling two of the rms spectra shown in Fig.
\ref{fig:grid}(b) to their peak value and plotting them on a linear
scale in Fig. \ref{fig:pabs}(a). The shape changes from being dominated
by a peak at $\sim$1~keV (lowest column) to being dominated by a peak
at $\sim$2~keV (highest column) with intermediate columns showing both
peaks in varying ratios. This is due to differential variability of the
two major absorption systems at $\sim$0.9~keV and 1.2--2~keV. The
former is produced predominantly by lines while the latter is
predominantly edges (Fig. \ref{fig:pabs}b). The lines easily become
optically thick at high columns, so their effect on the spectrum
saturates, and the rms spectra depend more on the opacity of the edges.
High ionization iron L edges dominate for $\langle \lg\xi \rangle$ =
3.0, giving rise to the peak at $\sim$2 keV. At $\langle \lg\xi
\rangle$ = 2.6 lower ionization iron ions dominate, so the peak at high
column is at $\sim $1.2 keV (Fig. \ref{fig:grid}a). The point at which
the resonance lines become optically thick (and hence insensitive to
changes in ionization) depends on the turbulent velocity assumed for
the absorption models. We assumed 100~km s$^{-1}$ but the velocity
shear required for the smearing of 0.2--0.3~$c$ may be a more realistic
estimate for this. This would lead to double peaked rms spectra over a
wider range of parameters.

We also explored the effect of an alternative variability pattern,
driven by changes in $N_H$ rather than ionization. The effect of this
(with constant ionization) is generally to change the depth of all the
absorption features together. Thus the rms spectra produced by this
variability mode are generally single-peaked.

\subsection{Application to the data}
\label{sec:data}

\begin{figure*}
\begin{center}
\leavevmode \epsfxsize=13cm \epsfbox{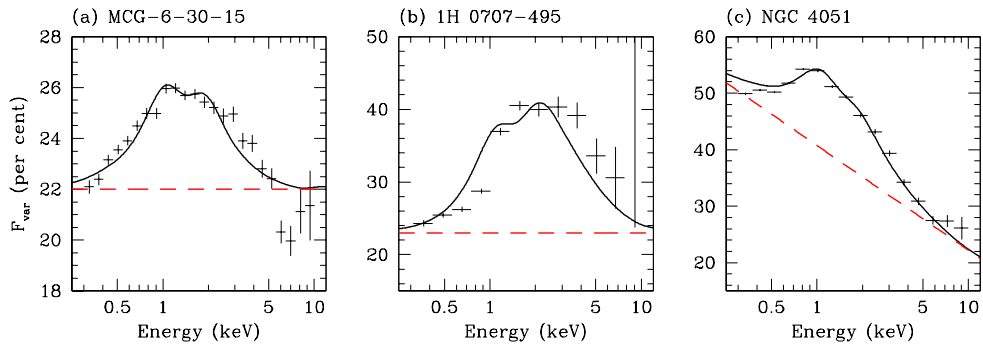}
\end{center}
\caption{The {\it XMM-Newton} rms spectra of MCG--6-30-15, 1H~0707--495
and NGC~4051 (crosses). The solid curves show the model of variable
ionization of the smeared absorber, roughly matched (not formally
fitted) to represent the data. In all cases the underlying power law
has index $\Gamma$ = 2.3 and the smearing is $\sigma_{\rm vel}$ = 0.3.
The ionization, column and variability are: (a) $N_H = 5\times10^{22}$
cm$^{-2}$, $\langle \lg \xi \rangle$ = 3.0 and $r(\lg \xi)$ = 0.036;
(b) $N_H = 2\times10^{23}$ cm$^{-2}$, $\langle \lg \xi \rangle$ = 3.5
and $r(\lg \xi)$ = 0.08; (c) $N_H = 5\times10^{22}$ cm$^{-2}$, $\langle
\lg \xi \rangle$ = 3.1 and $r(\lg \xi)$ = 0.12. The dashed lines show
variability of the continuum: (a) and (b) the energy-independent
component (constant rms); (c) the pivoting power law (about 160 keV). }

\label{fig:3agn}
\end{figure*}

Encouraged by these results we have attempted to apply the models
presented in Sec. \ref{sec:models} to observations of AGN. We have
calculated variability spectra from {\it XMM-Newton} data of NGC 4051
(2001 May 16/17 observation), MCG--6-30-15 (2001 August 2/3) and
1H~0707--495 (2002 October 13/14). We used background subtracted light
curves with 256-s resolution, accumulated from all available
instruments (PN only for NGC~4051, both MOS and PN for MCG--6-30-15 and
1H 0707--495). We show these spectra as crosses in Fig. \ref{fig:3agn}.

The rms spectra have rather broad peaks, indicating that models with
contributions from both absorption systems may be favoured by the data.
Accordingly, we have used the models where the variability is produced
just from changes in the ionization parameter, as in Figs.
\ref{fig:varrms} and \ref{fig:grid}.  We assumed that this was driven
by the power-law continuum variability (with spectral index of $\Gamma$
= 2.3), which produced an additional contribution to the rms.
Normalization changes of the power law produce an energy-independent
term (i.e. constant rms) to the variability spectrum (Gierli{\'n}ski \&
Zdziarski 2005). When adding power law and absorber variabilities, we
assumed that they were correlated and therefore the total standard
deviation was the sum of component standard deviations. We have not
performed formal spectral fits, but only roughly matched the model to
the data.

Such a composite model of the variability can describe the rms spectra
seen from MCG--6-30-15 and 1H 0707--495, as shown by solid curves in
Figs. \ref{fig:3agn}(a) and (b). The dashed lines correspond to the
energy-independent component due to varying power law. However, the
variability spectra from NGC~4051 could not be described in this way.
The spectrum shown in Fig.~\ref{fig:3agn}(c) appears to have the broad
peak between 0.8--2~keV superimposed not on a constant rms background
but on one which has more variability at lower energies. This
background rms can be modelled by a power law which varies in both
normalization and spectral index. We found that a power law pivoting
about an energy of 160 keV (see also Uttley et al. 2004) added to the
smeared absorber variability matches the data well (Fig.
\ref{fig:3agn}c). Such continuum variability has been seen in Galactic
black holes in the low/hard spectral state (Gierli{\'n}ski \& Zdziarski
2005).

Clearly, such a simple model can only be an approximation. The energy
spectra of AGN are complex and require multiple additional components.
For example, 1H 0707--495 shows a large drop in the energy spectrum
above $\sim$7 keV (e.g. Fabian et al. 2004), which cannot be explained
by a highly ionized ($\lg \xi \sim 3.5$) absorber in our model. This
feature requires a more complex model of the velocity structure than
the simple Gaussian smearing assumed here, if it arises from the
absorber. Additional spectral components can also respond in some way
to the changing X-ray illumination and add to the rms spectra. This may
account for the features in the rms spectra not covered by our model.
In particular, there is a dip in the rms around 7 keV in MCG--6-30-15,
not covered by our model. This could be due to a reflected component
which remains constant while the power law continuum varies (e.g.
Vaughan \& Fabian 2004).  Similarly, a few rms spectra in the
literature do not show the characteristic broad peak between 1--2~keV
e.g. NGC 4051 (November 2002 observation; Ponti et al. 2006), NGC 4151
(Schurch \& Warwick 2002) and I Zw 1 (Gallo et al. 2004c) show quite
different rms spectra, with a sudden drop in rms below $\sim$2--3 keV.
This kind of a variability spectrum cannot be reproduced by our simple
variability model.

If ionization variability is driven by changes in the central source
luminosity, then we expect the fractional variability amplitudes of
ionization and luminosity to be similar. However, this only holds to
within a factor $\sim$3. From the best-matching logarithm ionization
variability, $r(\lg \xi)$, we can find the corresponding ionization rms
amplitude, $r(\xi)$ = 25, 60 and 76 per cent, for MCG--6-30-15, 1H
0707--495 and NGC~4051, respectively. Our best estimate for the
power-law variability (including the pivoting of the spectral index in
NGC~4051) yields only about 22, 23 and 27 per cent variability in
luminosity, respectively. It is difficult to quantify the statistical
uncertainties in these numbers, but if the level of variability of the
ionization and continuum are significantly different then this could
indicate that the model rms spectra are somewhat distorted due to the
assumed turbulent velocity (see Section \ref{sec:models}). Another
possibility is that there are other components (which are known to be
present in the spectrum) contributing (or suppressing) the continuum
variability.

The aim of this Letter is to demonstrate that the model in which
changes in the smeared absorption driven be the changing continuum can
match well typical energy-dependent variability. It neither can
describe all the variability seen, nor it can account for all the
components required in spectral fitting. Modelling the effect of these
is beyond the scope of this short Letter, but we plan to include this
in later work.

\subsection{Properties of the smeared absorber}
\label{sec:absorber}

The smeared warm absorber model for the soft excess requires
substantial columns (few to few tens of $10^{22}$~cm$^{-2}$) with
$\lg\xi$ = 3.0--3.5 and substantial velocity shear ($\sim$0.3~$c$) in
order to fit typical AGN spectra (GD04; Chevallier et al. 2006; Schurch
\& Done in press; SD06).  Here we have shown that similar parameters
can also give an excellent description of the typical rms variability
spectra seen from these objects. However, these parameters are quite
extreme, so we examine the physical requirements in more detail.

Chevallier et al. (2006) show from the observed $N_H$ and $\xi$ that if
the material is a wind then it must have very small volume filling
factor, i.e. must consist of small dense clumps. Schurch \& Done (in
press) invert this argument to show instead that if the material has
large filling factor then it is not a wind. Instead they suggest that
it has a turbulent velocity structure where the material circulates (a
failed wind) rather than escaping to infinity. Either of these
situations might arise in the complex, high velocity accretion flow
structures close to the central engine and base of the jet (e.g. Hawley
\& Krolik 2006). Assuming that the material is a few tens of
Schwarzschild radii from a black hole of mass $M$, accreting at half
Eddington, the observed ionization parameter of $\xi\sim 10^3$ erg cm
s$^{-1}$ requires that the density $n=L/\xi R^2$ is of order $10^{21}
($M$_\odot/M$) cm$^{-3}$ = $10^{13-15}$~cm$^{- 3}$ for a black hole of
mass $10^{8-6} $M$_\odot$, respectively. Such high densities mean that
the ionization state of the material will change very rapidly in
response to changes in the continuum flux. There are no light travel
time delays as the absorption is along the line of sight, so the
ionization response to continuum changes occur on the recombination
timescales of the absorber, $T_{\rm recomb}\sim2.5\times10^{10}$
(cm$^{-3}$/$n$) s for {\sc oviii} (e.g. M$^c$Hardy et al. 1995). The
shortest continuum variability time-scales in these sources are all
$\sim$1000--2000 s (comparable with the dynamical time scales in the
innermost part of the disc) while the recombination timescale is only a
fraction of a second.

Hence, the model requirement that the ionization state can change in
response to the rapid continuum changes is consistent with the
properties of the material already inferred from the spectral fits.
Ionization changes are a necessary outcome of changing the illumination
of such material. Partially ionized absorption with a large velocity
shear can give a self-consistent explanation of both the energy spectra
and rms variability spectra.

\section{Conclusions}
\label{sec:discussion}

A crucial property of the soft X-ray excess in AGN is constancy of its
temperature when modelled by a thermal component (Czerny et al. 2003;
GD04; Crummy et al. 2005), which can be naturally explained if the
excess originates from atomic processes. One particular scenario
involves ionized smeared reflection (Crummy et al. 2006) from the
accretion disc, which can successfully fit the X-ray spectra (e.g.
Fabian et al. 2002b; Fabian et al. 2005). It has been shown recently
that the same model can match the rms spectra as well (Ponti et al.
2006). An alternative solution requires relativistically smeared,
partially ionized absorption (GD04). This model gives comparably good
fits to the X-ray spectra (SD06). In this paper we show for the first
time that the absorption model gives a very good description of the
shape of the 0.3--10~keV rms spectra, assuming that the continuum
variability drives correlated changes in the ionization parameter of
the smeared absorption.

\section*{Acknowledgements}

We thank Tim Kallman for help with {\sc xstar}. MG and CD acknowledge
support through a UK Particle Physics and Astronomy Research Council
(PPARC) Postdoctoral Research Fellowship and Senior Fellowship,
respectively.

\label{lastpage}


\begin{thebibliography}{}

\bibitem[\protect\citeauthoryear{Bautista \&
Kallman}{2001}]{2001ApJS..134..139B} Bautista M.~A., Kallman T.~R.,
2001, ApJS, 134, 139

\bibitem[\protect\citeauthoryear{Chevallier et
al.}{2006}]{2006A&A...449..493C} Chevallier L., Collin S., Dumont
A.-M., Czerny B., Mouchet M., Gon{\c c}alves A.~C., Goosmann R., 2006,
A\&A, 449, 493

\bibitem[\protect\citeauthoryear{Crummy et al.}{2006}]{2006MNRAS.365.1067C}
Crummy J., Fabian A.~C., Gallo L., Ross R.~R., 2006, MNRAS, 365, 1067

\bibitem[\protect\citeauthoryear{Czerny et al.}{2003}]{2003A&A...412..317C}
Czerny B., Niko{\l}ajuk M., R{\'o}{\.z}a{\'n}ska A., Dumont A.-M.,
Loska Z., Zycki P.~T., 2003, A\&A, 412, 317

\bibitem[\protect\citeauthoryear{Edelson et
al.}{2002}]{2002ApJ...568..610E} Edelson R., Turner T.~J., Pounds K.,
Vaughan S., Markowitz A., Marshall H., Dobbie P., Warwick R., 2002,
ApJ, 568, 610

\bibitem[\protect\citeauthoryear{Fabian \&
Miniutti}{2005}]{2005astro.ph..7409F} Fabian A.~C., Miniutti G., in
Wiltshire D. L., Visser M., Scott S. M., eds, Kerr Spacetime: Rotating
Black Holes in General Relativity. Cambridge Univ. Press, Cambridge, in
press (astro-ph/0507409)

\bibitem[\protect\citeauthoryear{Fabian et al.}{2002a}]{2002MNRAS.331L..35F}
Fabian A.~C., Ballantyne D.~R., Merloni A., Vaughan S., Iwasawa K.,
Boller T., 2002a, MNRAS, 331, L35 

\bibitem[\protect\citeauthoryear{Fabian et al.}{2002b}]{2002MNRAS.335L...1F}
Fabian A.~C., et al., 2002b, MNRAS, 335, L1 

\bibitem[\protect\citeauthoryear{Fabian et al.}{2004}]{2004MNRAS.353.1071F}
Fabian A.~C., Miniutti G., Gallo L., Boller T., Tanaka Y., Vaughan S.,
Ross R.~R., 2004, MNRAS, 353, 1071

\bibitem[\protect\citeauthoryear{Fabian et al.}{2005}]{2005MNRAS.361..795F}
Fabian A.~C., Miniutti G., Iwasawa K., Ross R.~R., 2005, MNRAS, 361,
795

\bibitem[\protect\citeauthoryear{Gallo et al.}{2004}]{2004MNRAS.347..269G}
Gallo L.~C., Boller T., Tanaka Y., Fabian A.~C., Brandt W.~N., Welsh
W.~F., Anabuki N., Haba Y., 2004a, MNRAS, 347, 269 

\bibitem[\protect\citeauthoryear{Gallo et al.}{2004}]{2004MNRAS.353.1064G}
Gallo L.~C., Tanaka Y., Boller T., Fabian A.~C., Vaughan S., Brandt
W.~N., 2004b, MNRAS, 353, 1064 

\bibitem[\protect\citeauthoryear{Gallo et al.}{2004}]{2004A&A...417...29G}
Gallo L.~C., Boller T., Brandt W.~N., Fabian A.~C., Vaughan S., 2004c,
A\&A, 417, 29 

\bibitem[\protect\citeauthoryear{Gierli{\'n}ski \&
Done}{2004}]{2004MNRAS.349L...7G} Gierli{\'n}ski M., Done C., 2004,
MNRAS, 349, L7 (GD04)

\bibitem[\protect\citeauthoryear{Gierli{\'n}ski \&
Zdziarski}{2005}]{2005MNRAS.363.1349G} Gierli{\'n}ski M., Zdziarski
A.~A., 2005, MNRAS, 363, 1349

\bibitem[\protect\citeauthoryear{Green, McHardy, \&
Done}{1999}]{1999MNRAS.305..309G} Green A.~R., McHardy I.~M., Done C.,
1999, MNRAS, 305, 309

\bibitem[\protect\citeauthoryear{Grupe et al.}{2004}]{2004AJ....127..156G}
Grupe D., Wills B.~J., Leighly K.~M., Meusinger H., 2004, AJ, 127, 156

\bibitem[\protect\citeauthoryear{Hawley \&
Krolik}{2006}]{2006ApJ...641..103H} Hawley J.~F., Krolik J.~H., 2006,
ApJ, 641, 103

\bibitem[\protect\citeauthoryear{Inoue \&
Matsumoto}{2003}]{2003PASJ...55..625I} Inoue H., Matsumoto C., 2003,
PASJ, 55, 625

\bibitem[\protect\citeauthoryear{Kinkhabwala}{2003}]{2003PhDT.........8K}
Kinkhabwala A.~A., 2003, PhDT

\bibitem[\protect\citeauthoryear{McHardy et
al.}{1995}]{1995MNRAS.273..549M} M$^c$Hardy I.~M., Green A.~R., Done
C., Puchnarewicz E.~M., Mason K.~O., Branduardi-Raymont G., Jones
M.~H., 1995, MNRAS, 273, 549

\bibitem[\protect\citeauthoryear{Magdziarz et
al.}{1998}]{1998MNRAS.301..179M} Magdziarz P., Blaes O.~M., Zdziarski
A.~A., Johnson W.~N., Smith D.~A., 1998, MNRAS, 301, 179

\bibitem[\protect\citeauthoryear{Markowitz, Edelson, \&
Vaughan}{2003}]{2003ApJ...598..935M} Markowitz A., Edelson R., Vaughan
S., 2003, ApJ, 598, 935

\bibitem[\protect\citeauthoryear{Miniutti \&
Fabian}{2004}]{2004MNRAS.349.1435M} Miniutti G., Fabian A.~C., 2004,
MNRAS, 349, 1435

\bibitem[\protect\citeauthoryear{Miniutti et
al.}{2003}]{2003MNRAS.344L..22M} Miniutti G., Fabian A.~C., Goyder R.,
Lasenby A.~N., 2003, MNRAS, 344, L22

\bibitem[\protect\citeauthoryear{Ponti et al.}{2004}]{2004A&A...417..451P}
Ponti G., Cappi M., Dadina M., Malaguti G., 2004, A\&A, 417, 451

\bibitem[\protect\citeauthoryear{Ponti et al.}{2006}]{2006MNRAS.368..903P}
Ponti G., Miniutti G., Cappi M., Maraschi L., Fabian A.~C., Iwasawa K.,
2006, MNRAS, 368, 903

\bibitem[\protect\citeauthoryear{Schurch \&
Warwick}{2002}]{2002MNRAS.334..811S} Schurch N.~J., Warwick R.~S.,
2002, MNRAS, 334, 811

\bibitem[]{}Schurch N. J., Done C., MNRAS, in press

\bibitem[\protect\citeauthoryear{Turner \&
Pounds}{1988}]{1988MNRAS.232..463T} Turner T.~J., Pounds K.~A., 1988,
MNRAS, 232, 463

\bibitem[\protect\citeauthoryear{Uttley et al.}{2004}]{2004MNRAS.347.1345U}
Uttley P., Taylor R.~D., M$^{\rm c}$Hardy I.~M., Page M.~J., Mason
K.~O., Lamer G., Fruscione A., 2004, MNRAS, 347, 1345

\bibitem[\protect\citeauthoryear{Vaughan \&
Fabian}{2004}]{2004MNRAS.348.1415V} Vaughan S., Fabian A.~C., 2004,
MNRAS, 348, 1415

\bibitem[\protect\citeauthoryear{Vaughan et
al.}{2003}]{2003MNRAS.345.1271V} Vaughan S., Edelson R., Warwick R.~S.,
Uttley P., 2003, MNRAS, 345, 1271

\bibitem[\protect\citeauthoryear{Wilms et al.}{2001}]{2001MNRAS.328L..27W}
Wilms J., Reynolds C.~S., Begelman M.~C., Reeves J., Molendi S.,
Staubert R., Kendziorra E., 2001, MNRAS, 328, L27

\bibitem[\protect\citeauthoryear{Young et al.}{2005}]{2005ApJ...631..733Y}
Young A.~J., Lee J.~C., Fabian A.~C., Reynolds C.~S., Gibson R.~R.,
Canizares C.~R., 2005, ApJ, 631, 733

\end{thebibliography}
\end{document}